\pdfoutput=1
\documentclass[pra,twocolumn,superscriptaddress,showpacs,floatfix]{revtex4}

\usepackage{graphics,epsf,times,amsfonts,bm,bbm}
\usepackage{latexsym}
\usepackage{amsthm}
\usepackage{amssymb}
\usepackage{amsmath}
\usepackage{graphicx}
\usepackage[
colorlinks=true,linkcolor=red,citecolor=green,urlcolor=blue,
pdfstartview=FitH,
pagebackref=false,
pdfauthor={Jacob R. West, Daniel A. Lidar, Bryan H. Fong, Mark F. Gyure}, 
pdftitle={High fidelity quantum gates via dynamical decoupling},
pdfsubject={quantum information processing}, 
pdfkeywords={quantum, information, computation, dynamical decoupling}]{hyperref}
\usepackage{hypernat}

\DeclareMathOperator{\DD}{DD}
\DeclareMathOperator{\CDD}{CDD}
\DeclareMathOperator{\PDD}{PDD}

\newcommand{\ket}[1]{\left\vert#1\right\rangle}

\begin{document}

\title{High fidelity quantum gates via dynamical decoupling}
\author{Jacob R. \surname{West}}
\affiliation{HRL Laboratories, LLC., 3011 Malibu Canyon Rd., Malibu, California 90265, USA}
\author{Daniel A. \surname{Lidar}}
\affiliation{Departments of Electrical Engineering, Chemistry, and Physics, Center for
Quantum Information \& Technology, University of Southern California, Los
Angeles, California 90089, USA}
\author{Bryan H. \surname{Fong}}
\affiliation{HRL Laboratories, LLC., 3011 Malibu Canyon Rd., Malibu, California 90265, USA}
\author{Mark F. \surname{Gyure}}
\affiliation{HRL Laboratories, LLC., 3011 Malibu Canyon Rd., Malibu, California 90265, USA}
\date{\today}

\begin{abstract}
Realizing the theoretical promise of quantum computers will require
overcoming decoherence. Here we demonstrate numerically that high fidelity
quantum gates are possible within a framework of quantum dynamical
decoupling. Orders of magnitude improvement in the fidelities of a universal
set of quantum gates, relative to unprotected evolution, is achieved over a
broad range of system-environment coupling strengths, using recursively
constructed (concatenated) dynamical decoupling pulse sequences.
\end{abstract}

\maketitle

\textit{Introduction.---}Quantum systems are famously susceptible to
interactions with their surrounding environments, a process which leads to a
progressive loss of \textquotedblleft quantumness\textquotedblright\ of
these systems, via decoherence \cite{Zurek}. When a system performs a
quantum information processing (QIP) task this loss of quantumness is
equivalent to the accumulation of computational errors, which leads to the
eventual loss of any quantum advantage in information processing. Robust
large-scale quantum information processing therefore requires that
decoherence---or any otherwise undesired evolution---of a quantum state be
minimized to the largest extent possible by clever system choice and
engineering. One may then hope to apply the powerful techniques of fault
tolerant quantum error correction (FT-QEC) \cite{Gaitan:book}. However,
FT-QEC imposes significant resource requirements, in particular rapidly
growing spatial and temporal overhead, together with demanding gate and
memory error rates which must remain below a certain threshold (e.g., Refs.~%
\cite{Aliferis:05Knill:05}). This motivates the search for alternative
strategies which can slow down decoherence and \textquotedblleft keep
quantumness alive.\textquotedblright\ Dynamical decoupling (DD) is a form of
quantum error \emph{suppression} that modifies the system-environment
interaction so that its overall effects are very nearly self-canceling,
thereby decoupling the system evolution from that of the noise-inducing
environment \cite{Viola:99Zanardi:98b}. DD has primarily been studied
as a specialized control technique for quantum memory (i.e., arbitrary
state preservation)
\cite{Viola:05,KhodjastehLidar:04KhodjastehLidar:07,Witzel:07a,Zhang:08,Uhrig:07,yang:180403,WFL:09,UL:10},
as convincingly demonstrated by a number of recent experiments in QIP
platforms as diverse as electron-nuclear systems \cite{li:190401Morton:08,Du:09}, photonics qubits \cite{Damodarakurup:08}, and
trapped ions \cite{Biercuk:09}. However, the holy grail of QIP is not
just to store states robustly, but rather to perform \emph{universal
computation} robustly \cite{Nielsen:book}. Fortunately, there are abstract
results showing that DD is in principle compatible with computation,
essentially by designing DD\ operations that commute with the computational
operations \cite{Viola:99a}. Additionally, recent theoretical results
indicate that high fidelity \textquotedblleft dynamically error-corrected
gates\textquotedblright\ can be designed, using methods inspired by DD \cite%
{khodjasteh:080501KLV:09}, and that DD can be merged with FT-QEC to reduce
resource overhead, or even improve gate error rates to below threshold \cite%
{NLP:09}. DD can also be used to improve the fidelity of adiabatic quantum
computation \cite{Lidar:AQC-DD}. Experimentally, DD has been successfully
combined with QEC in nuclear spin systems to demonstrate robust quantum
memory \cite{Boulant:02}. In principle, then, it appears that DD is a
suitable control technique for overcoming decoherence and improving gate
fidelity. However, a very practical question still remains: what are the
conditions under which DD can be used to perform universal quantum
computation with a given fidelity? Recent rigorous bounds devised for
the popular ``periodic DD'' (PDD) protocol suggest that it is severely
limited in this regard \cite{KhodjastehLidar:08}. Here we demonstrate, using numerical
simulations of a logical qubit coupled to a small bath, that recursively
constructed, concatenated DD (CDD) pulse sequences
\cite{KhodjastehLidar:04KhodjastehLidar:07}, can be used to endow a
universal set 
of quantum logic gates with remarkably high fidelities. Though our numerical
results illustrate the effectiveness of CDD in a model inspired by quantum
dot implementations of QIP \cite{Burkard:99}, the framework we describe is in principle more
generally applicable, provided encoding and encoded gates can be
implemented. We therefore expect that our results will contribute to
establishing DD as an indispensable tool in scalable QIP.

\textit{Dynamical decoupling.---}The total Hamiltonian without DD\ is $%
H=H_{B}+H_{SB}$, where $H_{B}$ includes all bath-only terms and $H_{SB}$
includes all terms acting non-trivially on the system. To suppress error, DD
allows the joint evolution to proceed under $H$ for some time before
applying a control pulse $P_{j}$ to the system alone [generated by a
time-dependent system-only Hamiltonian $H_{S}(t)$ which is added to $H$],
designed to refocus the evolution toward the error-free ideal, continually
repeating this process until some total evolution has been completed: 
$\DD[U(\tau_0)]=P_{N}U(\tau _{0})\cdots P_{2}U(\tau _{0})P_{1}U(\tau
_{0})\equiv \widetilde{U}(N\tau _{0})$, where $U(\tau _{0})=U_{0}(\tau
_{0})B(\tau _{0})$
represents the joint system-bath unitary evolution generated by $H$, for a
duration of length $\tau _{0}$ (the pulse interval), decomposed so that $%
U_{0}(\tau _{0})$ determines the ideal, desired system-only error-free
evolution, and $B(\tau _{0})$ is a unitary error operator acting jointly on
the system and bath. Here and below we use a tilde to denote evolution
in the presence of DD pulses. DD schemes with non-uniform pulse intervals have also
been considered and shown to be very powerful for quantum memory purposes 
\cite{Uhrig:07,yang:180403,WFL:09,UL:10,Biercuk:09,Du:09}, but since it is unclear
how to use them for computation involving multiple qubits, we shall limit
our discussion to uniform-interval schemes. For now, but not in our
simulations presented below, we assume for simplicity of presentation that
the pulses $\{P_{j}\}$ are sufficiently fast as to not contribute to the
total time of the evolution. The simplest example is quantum memory, where $%
U_{0}(\tau _{0})=I_{S}$ is the identity operation,\ and $B(\tau _{0})$
represents the deviation from the ideal dynamics caused by the presence of a
bath. In this case, our goal is to choose pulses so that $\DD[U(\tau_0)]%
=I_{S}\otimes \widetilde{B}$, where $\widetilde{B}$ is an arbitrary
pure-bath operator. Uniform-interval DD schemes differ in precisely
how the 
pulses $\{P_{j}\}$ are chosen, with the only common constraint that the
following basic \textquotedblleft decoupling
condition\textquotedblright\ (vanishing average Hamiltonian, i.e.,
vanishing first order term in the Magnus series of the joint system-bath evolution) is
met \cite{Viola:99Zanardi:98b}: $\sum_{\alpha }P_{\alpha }^{\dag
}H_{SB}P_{\alpha }=0$. To be concrete, we will suppose that the pulses $%
P_{\alpha }\in \{I,X,Y,Z\}$ are Pauli operators. CDD generates pulse
sequences by recursively building on a base sequence $Z[\cdot ]X[\cdot
]Z[\cdot ]X[\cdot ]$ (motivated below), where $[\cdot ]$ denotes either free evolution
or the insertion of gate operations between pulses. The sequence is initialized as $\CDD%
_{0}[U(\tau _{0})]=U(\tau _{0})=U_{0}(\tau _{0})B(\tau _{0})\equiv 
\widetilde{U}_{0}(\tau _{0})$, and higher levels are generated via the rule $%
\CDD_{n+1}[U(\tau _{0})]=Z[\widetilde{U}_{n}(\tau _{n})]X[\widetilde{U}%
_{n}(\tau _{n})]Z[\widetilde{U}_{n}(\tau _{n})]X[\widetilde{U}_{n}(\tau
_{n})]\equiv \widetilde{U}_{n+1}(\tau _{n+1})$, where $\tau _{n}=4^{n}\tau
_{0}$. Note that in contrast to previous work on CDD \cite%
{KhodjastehLidar:04KhodjastehLidar:07,Witzel:07a,NLP:09}, we are allowing
for the possibility of some non-trivial information processing operation $%
U_{0}(\tau _{0})$, as this will be required in our discussion of universal
computation below. The choice of the base sequence is motivated by the
observation that it satisfies the \textquotedblleft decoupling
condition\textquotedblright , in the quantum memory setting $U_{0}(\tau
_{0})=I_{S}$, under the dominant \textquotedblleft
1-local\textquotedblright\ system-bath coupling term $H_{SB}^{(1)}=\sum_{%
\alpha \in \{x,y,z\}}\sum_{j}\sigma _{j}^{\alpha }\otimes B_{j}^{\alpha }$,
where $\sigma _{j}^{x}\equiv X$, $\sigma _{j}^{y}\equiv Y$, and $\sigma
_{j}^{z}\equiv Z$ denote the Pauli matrices acting on system qubit $j$, and $%
\{B_{j}^{\alpha }\}$ are arbitrary bath operators. (The next order
\textquotedblleft 2-local\textquotedblright\ coupling would have terms such
as $\sigma _{j}^{\alpha }\sigma _{k}^{\beta }\otimes B_{jk}^{\alpha \beta }$%
, etc.) Similarly, the
most common pulse sequence used thus far in DD experiments (e.g.,
Refs.~\cite{Boulant:02,li:190401Morton:08}) is PDD, which generates
pulse sequences 
by \emph{periodically} repeating the base sequence $%
Z[\cdot ]X[\cdot ]Z[\cdot ]X[\cdot ]$: $\PDD_{k}[U(\tau _{0})]=(\PDD%
_{1}[U(\tau _{0})])^{k}=\widetilde{U}_{k}(4k\tau _{0})$, where $\PDD%
_{1}[U(\tau _{0})]=\CDD_{1}[U(\tau _{0})]$. Rigorous noise reduction bounds
are known for both PDD and CDD in the quantum memory setting, and show that
CDD is a much more effective strategy than PDD, provided $(\left\Vert
H_{B}\right\Vert +\left\Vert H_{SB}\right\Vert )\tau _{0}$ is sufficiently
small, where the norm is the largest eigenvalue \cite{NLP:09}. It is convenient to characterize the
leading-order DD behavior in terms of $\left\Vert H_{B}\right\Vert $ and $%
\left\Vert H_{SB}\right\Vert $, as these parameters capture the strength or
overall rate of the internal bath and system-bath dynamics, respectively. If 
$\left\Vert H_{SB}\right\Vert \gg \left\Vert H_{B}\right\Vert $, then the
system-bath coupling is a dominant source of error. In this
case, DD should produce significant fidelity gains as it removes the
dominant error source. On the other hand, if $\left\Vert H_{SB}\right\Vert
<\left\Vert H_{B}\right\Vert $, then the system-bath coupling
induces relatively slow dynamics, while the environment itself has fast
internal dynamics. In this case, suppressing the system-bath coupling will
have less of an effect on the overall dynamics, so it may be considered a
worst case scenario when assessing DD performance.

\textit{High fidelity universal quantum gates using CDD.---} Our main
goal in this work is to demonstrate
that we can generate a universal set of logic gates which is highly robust
in the presence of a decohering environment. As a model system we consider
electron spin qubits in semiconductor quantum dots \cite{Burkard:99}, which we study
numerically via full-quantum-state (sometimes called \textquotedblleft
numerically exact\textquotedblright ) simulations over a wide range of
system-bath coupling parameters. In such systems the dominant bath is
provided by the nuclear spins \cite{Witzel:07a}, and the interaction between system and
environment is described by a Heisenberg exchange Hamiltonian with
exponentially decreasing strength as a function of distance $d_{ij}$ between
system qubit $j$ and bath qubit $i$. Thus, we let $B_{j}^{\alpha
}=J\sum_{i}\sigma _{i}^{\alpha }/2^{d_{ij}}$ in the system-bath Hamiltonian $%
H_{SB}^{(1)}$, so that $\left\Vert H_{SB}\right\Vert \propto J$. We model
the interaction between the bath nuclear spin qubits as dipole-dipole
coupling, i.e., $H_{B}=\beta \sum_{i<j}\left( \sigma _{i}^{y}\sigma
_{j}^{y}+\sigma _{i}^{z}\sigma _{j}^{z}-2\sigma _{i}^{x}\sigma
_{j}^{x}\right) /d_{ij}^{3}$ so that $\left\Vert H_{B}\right\Vert \propto
\beta $, where now $d_{ij}$ is the distance between bath qubits $i$ and $j$.
In our simulations we pick the parameters $J$, $\beta $, and $d_{ij}$, as
well as the pulse interval $\tau_0$ and the pulse width, to include a range of interest for
GaAs and Si quantum dots \cite{Petta,supp-mat}. The $H_{SB}$ and $H_{B}$
Hamiltonians are on during the entire pulse sequence execution, while $%
H_{S}(t)$ pulses appropriately between dynamical decoupling ($H_S=H_{DD}$) and
computational operations ($H_S=H_G$).

Universal quantum computation requires that only a discrete set of universal
gates be implemented; a particularly simple choice are the Hadamard, $\pi /8$%
, and controlled-phase gates \cite{Nielsen:book}. The first two are
single-qubit gates, and the third is a two-qubit gate which can be used to
generate entanglement. A conundrum immediately presents itself when
trying to combine computation with DD: how to make sure that the DD pulses do not cancel the
(system Hamiltonian implementing the) gates? One solution is to use an
encoding so that the DD operations commute with the logical
gate operations \cite{Viola:99a,Lidar:AQC-DD,ByrdLidar:01a}. To this end we use logical qubits encoded
into a four-qubit decoherence-free subspace (DFS). The logical basis states
are the two orthonormal total spin-zero states of four spin-$1/2$ particles,
first described as a DFS in Ref.~\cite{Zanardi:97c}. We stress that our system-bath
interaction does not exhibit any symmetries so that there is no naturally
occuring DFS which can be used to store protected quantum information;
instead, our encoding choice is motivated by the fact that in this setting,
a universal set of encoded computational operations can be generated
by controllable
Heisenberg exchange Hamiltonians between the system qubits ($=H_G$), as first
described in \cite{Bacon:99aKempe:00}, and
these commute with the global Pauli operations $\left\{ \bar{X},\bar{Z}%
\right\} =\left\{ X_{1}X_{2}X_{3}X_{4},Z_{1}Z_{2}Z_{3}Z_{4}\right\} $ used
as decoupling pulses. To generate these pulses 
$H_{DD}$ is modeled as a controllable uniform magnetic field. However, we emphasize that these choices are by no
means unique. Any choice of DD pulses $\{P_{j}\}$ such that
$[H_G,P_{j}]=0$ $\forall j$ will suffice \cite{Viola:99a},
including, e.g., the stabilizer quantum error correcting codes relevant in
the theory of FT-QEC used as DD pulses, and the normalizers of these codes
used to generate computational gates \cite{ByrdLidar:01a,Lidar:AQC-DD}.

Faced with several options for combining DD and computational
operations \cite{Viola:99a,KhodjastehLidar:08}, we chose the following
\textquotedblleft decouple while compute\textquotedblright\
strategy. In this strategy we alternate between applying computational
and DD operations, thus spreading\ a computational gate over the
entire CDD pulse sequence. We do this by applying the
$N^{\mathrm{th}}$ root of the gate $N$ times during a CDD pulse
sequence involving $N=4^{n}$ pulses. Thus, if the ideal computational
gate is $G(T)=e^{-iH_{G}T}$, we implement it by applying $%
U(\tau _{0})=e^{-i(H_{G}+H_{SB}+H_{B})\tau _{0}}$ between each of the
$N$ pulses, where $\tau _{0}=T/N$.  As $T$ increases with
concatenation level $n$, the exchange couplings in $H_G$ are
proportionately decreased so that the
$\|H_G\| T$
product remains
constant; in all simulations the pulse interval $\tau_0$ is held fixed
at 1ns, the time-scale for exchange operations in semiconductor
quantum dots \cite{Petta}.  This ``decouple while compute'' strategy is
precisely the formulation presented in the expressions for
$\CDD_{n+1}[U(\tau _{0})]$ and $\PDD_{k}[U(\tau _{0})] $ above,
provided we identify $U_{0}(\tau _{0})B(\tau _{0})$ there with $U(\tau
_{0})$ here, and $U_{0}(\tau _{0})$ there with $G(\tau _{0})$
here. Other strategies are certainly also conceivable, e.g., a
\textquotedblleft decouple then compute\textquotedblright\ strategy
wherein $%
U_{0}(\tau _{0})$ is simply the identity operation, and the gate is
implemented at the end of the pulse sequence. While the latter
strategy was shown to be capable of reducing the resource requirements
of FT-QEC \cite%
{NLP:09}, we found in our simulations that we obtain a higher fidelity
when we use \textquotedblleft decouple while compute\textquotedblright
, because then time is not wasted\ on free evolution during the
intervals between pulses.

\begin{figure}[tbp]
\includegraphics[width=3.5in]{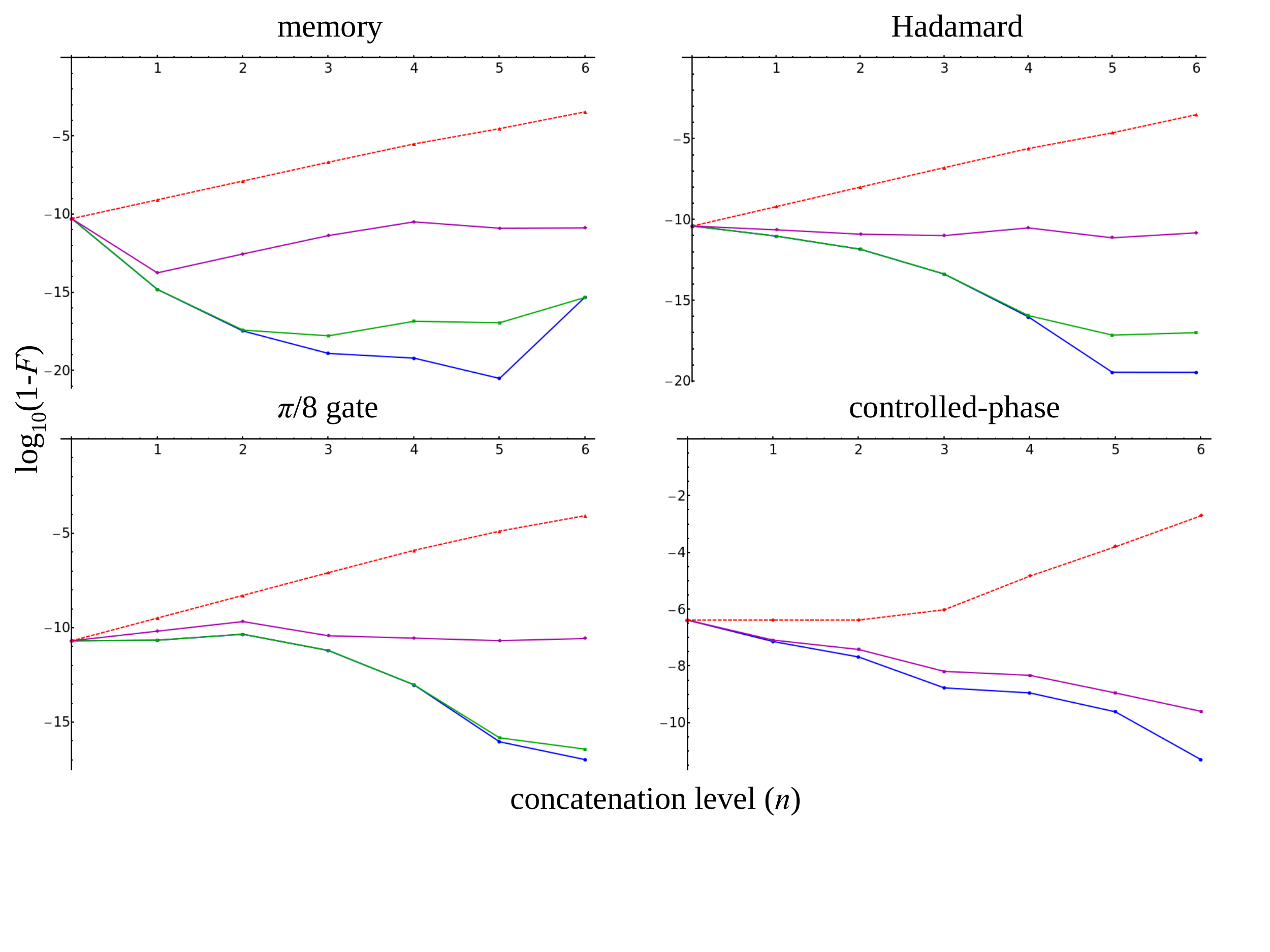}  \vspace{-1.5cm}
\caption{(color online) Fidelity of a universal set of encoded gates under CDD.
The coupling strengths and bath dynamics are determined by the parameters $%
J=10$kHz and $\protect\beta =1$MHz, respectively (we work in units of $\hbar
=1$). Pulse intervals are fixed at $\protect\tau _{0}=1$ns, while pulse
widths are given by $\protect\delta =0$, $\protect\delta =1$ps, and $\protect%
\delta =1$ns, corresponding, from bottom to top, to the blue, green (absent
for controlled-phase), and magenta lines, respectively. The red dashed line
shows the unprotected evolution over a time period $T=4^{n}\protect\tau _{0}$%
. Notice that the $\log _{10}(1-F)$ ranges change between plots. Also, $n=0$
corresponds to free evolution for a duration $\protect\tau _{0}=1$ns, whence
the $n=0$ point starts at a relatively high fidelity. Results depend
only slightly on the choice of initial system state.}
\label{ugatesfig}
\end{figure}

\begin{figure*}[tbp]
\includegraphics[width=6.5in]{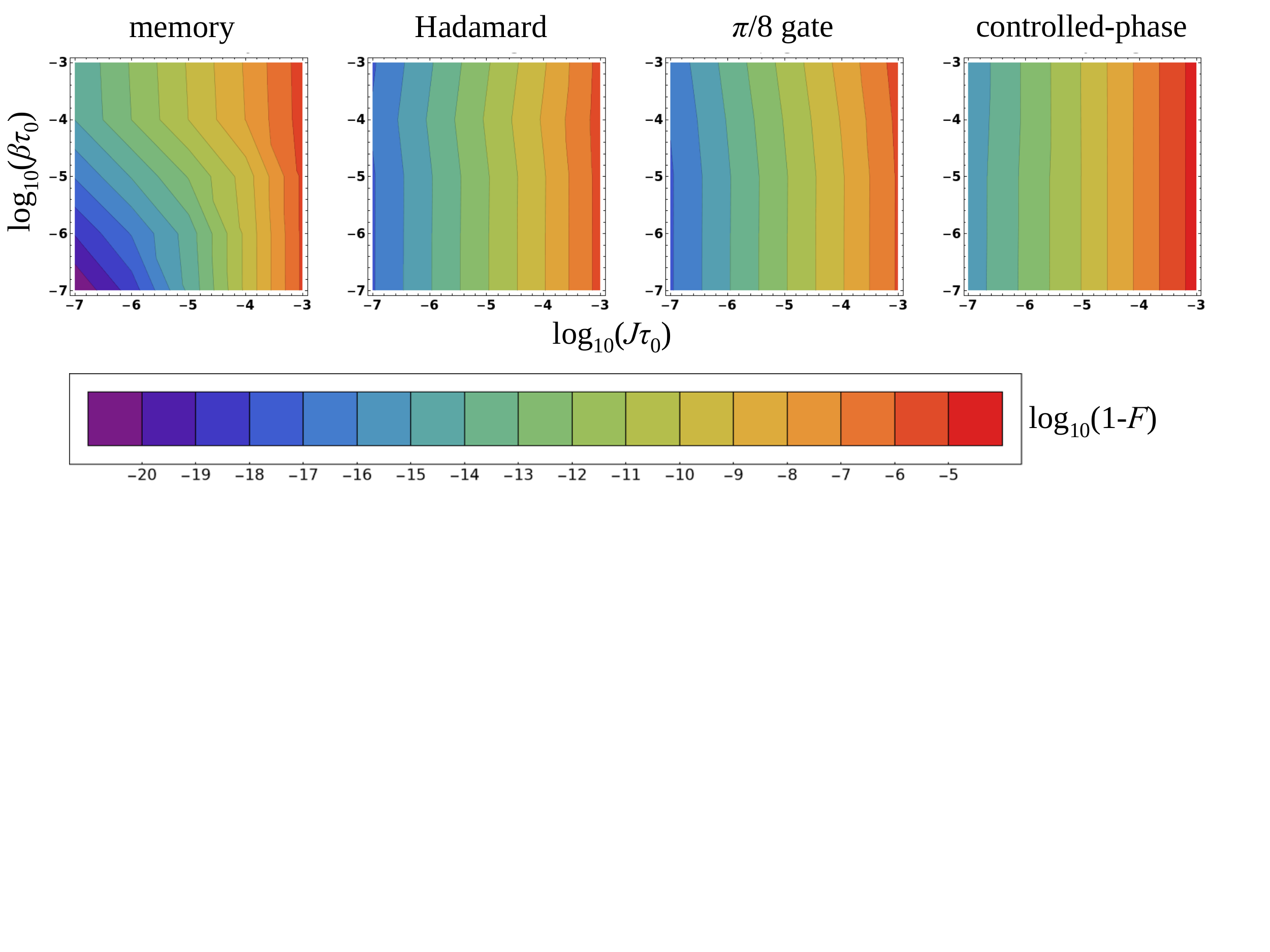} \vspace{-6.cm}
\caption{(color online) Constant fidelity contours for the system described in the previous
figure, at fixed concatenation level $n=5$ and pulse width $\protect\delta %
=1 $ns. Notice that the fidelity contours are strongly dependent on $J%
\protect\tau _{0}$.}
\label{contours}
\end{figure*}

We now present our simulation results (details of the numerical procedure
are given in \cite{supp-mat}). The worst-case scenario of $J<\beta $ is shown
in Fig.~\ref{ugatesfig} , where we plot $\log _{10}(1-F)$ vs. concatenation
level for each of the universal gates, with the fidelity defined as $F\equiv 
\sqrt{|\left\langle \psi \right\vert \rho \left\vert \psi \right\rangle |}$,
where $\rho $ is the mixed output system state (obtained from the joint
system-bath evolution after partial trace over the bath) and $\left\vert
\psi \right\rangle $ is the desired system state. In each of these plots,
the red dashed line represents undecoupled free evolution for increasing
total time, given by $T=4^{n}\tau _{0}$. As the evolution time increases,
error accumulates and fidelity correspondingly worsens, while $\CDD_{n}$
combats this effect with each successive level of concatenation. To
contrast, the blue line in these graphs shows $\CDD_{n}$ with ideal,
zero-width DD pulses, so that realistic, finite-width DD pulses lie
somewhere between the blue and red lines, as shown. In each of the plots in
Fig.~\ref{ugatesfig} CDD achieves impressive results, even when pulse
widths are as long as the intervals, that is, when $\delta =\tau _{0}=1$ns
as depicted with the magenta lines, CDD still manages more than five orders of
magnitude improvement in fidelity over free evolution. As the pulse width $%
\delta $ narrows relative to the pulse interval $\tau _{0}$, that is, as the
DD pulses becomes faster, fidelity improvement grows to between ten and
twenty orders of magnitude over free evolution.

The results for the encoded $\pi /8$ and Hadamard gates are similar, which
is not surprising given that they require, respectively, one and two
elementary Heisenberg exchange operations to be implemented \cite{Bacon:99aKempe:00,Bacon:thesisWoodworthMizelLidar:05}. The fidelity of the controlled-phase gate is several
orders of magnitude lower, which is due to the fact that it involves a much
longer sequence of $42$ elementary Heisenberg operations \cite%
{Bacon:thesisWoodworthMizelLidar:05}. Finally, while the quantum memory
results are comparable to those of the Hadamard and $\pi /8$-gates, we
attribute the reduction in memory fidelity at the highest concatenation
level to the absence of $H_{G}$ during the intervals between pulses. Indeed,
having the system Hamiltonian \textquotedblleft on\textquotedblright\ during
the pulse intervals has a beneficial effect, as it effectively reduces the
strength of the bath and system-bath Hamiltonians. The overall conclusion
from Fig.~\ref{ugatesfig} is rather encouraging: it appears to be possible to
implement a universal set of quantum logic gates with a high fidelity in the
presence of coupling to a spin bath.

The results in Fig.~\ref{ugatesfig} are for specific coupling parameters
chosen deliberately to represent a worst-case scenario for DD, in that $%
J<\beta $. As we next demonstrate, the conclusions are robust:\ CDD remains
effective over a broad range of bath dynamics and system-bath coupling
strengths. Figure~\ref{contours} shows the resilience of CDD to widely
varying environments by displaying constant fidelity contours in $(J\tau
_{0},\beta \tau _{0})$ space, at fixed concatenation level and pulse width,
as indicated. Note that fixing $n$ and $\delta $ renders the total evolution
time constant, so that fidelity becomes strictly a function of the
dimensionless coupling
parameters $(J\tau _{0},\beta \tau _{0})$. These plots show a strong
fidelity dependence on $J\tau _{0}$, and a very weak dependence on $\beta
\tau _{0}$, except in the quantum memory case.

More generally, our results show that CDD is effective over a broad range of
coupling parameters, including the fundamentally different \textquotedblleft
good\textquotedblright\ ($J>\beta $) and \textquotedblleft
bad\textquotedblright\ ($J<\beta $) regimes. This conclusion is further
bolstered by our complete gate fidelity simulations \cite{supp-mat}, where
the $\beta $ and $J$ parameters each vary over the range from $1$Hz to $1$%
MHz. In these simulations the non-memory gate fidelities improve
monotonically as a function of concatenation level for all values of $J$ and 
$\beta $. Taken in their totality, our simulation results indicate that
universal quantum computation can be combined with CDD to achieve very high
fidelities.

\textit{Discussion.---}The high gate fidelities we have reported here
suggest that it is advantageous to incorporate CDD as a first layer of defense
against decoherence, in a more complete FT-QEC scheme. In this regard our
\textquotedblleft decouple while compute\textquotedblright\ study
complements the \textquotedblleft decouple then compute\textquotedblright\
strategy for which it has already been shown that incorporating DD, and in
particular CDD, into FT-QEC can lead to substantial improvements \cite%
{NLP:09}. It will be interesting to determine which strategy leads to better
performance overall. Undoubtedly, optimal control methods will offer
important additional performance improvements in the implementation of
encoded logic gates \cite{cappellaro:044514}. We look
forward to experimental tests of CDD-protected quantum memory and
logic gates.

The views and conclusions contained in this document are those
of the authors and should not be interpreted as representing the official
policies, either expressly or implied, of the United States Department of
Defense or the U.S. Government.  Approved for public release,
distribution unlimited.

\textit{Acknowledgments}.--- All authors were sponsored by the
United States Department of Defense. D.A.L. was also sponsored by NSF
under Grants No. CHM-924318, CCF-726439 and PHY-803304.


\begin{widetext}
\appendix
\section{The DFS code}
\label{app:code}

We describe the four-qubit DFS code, first proposed 
in the context of providing immunity against collective decoherence
processes \cite{Zanardi:97c}. Let $S$ and $m_S$ denote the quantum numbers associated
with total spin and its projection, and let the singlet and
triplet states of two electrons $i,j$ be denoted as 
\begin{eqnarray*}
&& |s\rangle _{ij} \equiv |S=0,m_{S}=0\rangle =\frac{1}{\sqrt{2}}\left( |\Psi
(\uparrow \downarrow )\rangle -|\Psi (\downarrow \uparrow )\rangle \right) 
\\
&& |t_{-}\rangle _{ij} \equiv |S=1,m_{S}=-1\rangle =|\Psi (\downarrow
\downarrow )\rangle  \\
&& |t_{0}\rangle _{ij} \equiv |S=1,m_{S}=0\rangle =\frac{1}{\sqrt{2}}\left(
|\Psi (\uparrow \downarrow )\rangle -|\Psi (\downarrow \uparrow )\rangle
\right)  \\
&& |t_{+}\rangle _{ij} \equiv |S=1,m_{S}=1\rangle =|\Psi (\uparrow \uparrow
)\rangle .
\end{eqnarray*}
Here $|\Psi (\downarrow \uparrow )\rangle$ denotes a normalized basis
state with the first (second) electron in the spin up (down) state,
etc. Then a single encoded DFS qubit is formed by the two singlets of four spins,
i.e., the two states with zero total spin $S_{\mathrm{T}}=\left|
\mathbf{S}_{A}+\mathbf{S}_{B}+\mathbf{S}_{C}+\mathbf{S}_{D} \right|$,
where $\mathbf{S}_i$ is the Pauli spin vector-operator of electron $i$. These states are formed by combining two
singlets of two pairs of spins ($|0_{L}\rangle $), or triplets of two pairs
of spins ($|1_{L}\rangle $), with appropriate Clebsch-Gordan coefficients: 
\begin{eqnarray*}
|0_{L}\rangle  &=&|s\rangle _{AB}\otimes |s\rangle _{CD}  \notag \\
&=&\frac{1}{2}\left( |\Psi (\uparrow \downarrow \uparrow \downarrow )\rangle
+|\Psi (\downarrow \uparrow \downarrow \uparrow )\rangle -|\Psi (\uparrow \downarrow \downarrow \uparrow )\rangle
-|\Psi (\downarrow \uparrow \uparrow \downarrow )\rangle \right) 
\label{eq:0L} \\
|1_{L}\rangle  &=&\frac{1}{\sqrt{3}}\left( |t_{-}\rangle _{AB}\otimes
|t_{+}\rangle _{CD}-|t_{0}\rangle _{AB}\otimes |t_{0}\rangle _{CD}
+|t_{+}\rangle _{AB}\otimes |t_{-}\rangle _{CD}\right)  
\notag \\
&=&\frac{1}{\sqrt{3}}(2|\Psi (\uparrow \uparrow \downarrow \downarrow
)\rangle +2|\Psi (\downarrow \downarrow \uparrow \uparrow )\rangle -|\Psi
(\uparrow \downarrow \downarrow \uparrow )\rangle -|\Psi (\downarrow \uparrow \uparrow \downarrow )\rangle -|\Psi
(\uparrow \downarrow \uparrow \downarrow )\rangle -|\Psi (\downarrow
\uparrow \downarrow \uparrow )\rangle ).  \label{eq:1L}
\end{eqnarray*}

The details of the sequences of exchange interactions needed to
implement the Hadamard, $\pi/8$, and controlled-phase gates over this
DFS code are too long to give here. However, they have been well
documented in
the original works \cite{Bacon:99a,Kempe:00,Bacon:thesis}, and perhaps
most concisely in the more recent Ref.~\cite{WoodworthMizelLidar:05}. 

\section{Details of the numerical procedure}
\label{app:num}

We performed numerically exact simulations since we required extremely high precision fidelity
results, inaccessible via approximation
techniques \cite{witzel:035322,yang:085315} capable of handling much
larger system and bath sizes. Such methods are required for
long time scale simulations, which is not our case. Moreover,
large-scale simulations \cite{Witzel:07a} have confirmed earlier
small-bath simulations of CDD pulse sequences \cite{KhodjastehLidar:04,KhodjastehLidar:07}.
Our numerically exact simulations ran at 100-200 digits of numerical
precision, for more than 90 hours on a computer with a dual core intel
processor (2GHz + 2G RAM).  In our simulations, each logical qubit was
encoded using four physical qubits. We tested for dependence on the
initial encoded system state, and found a variation in output fidelity
of less than an order of magnitude. Hence all our reported results are
for the logical-one initial state.  As a further test we considered
several simple bath geometries: linear, circular, and polygonal. Our
results did not depend appreciably on this geometry.  We took the
initial bath state as the uniform superposition, zero temperature
state $\frac{1}{\sqrt{B}}\sum_{i=1}^B \ket{i}$, where $B = 2^{N_n}$ is
the number of available pure bath states when $N_n$ bath qubits are
present.  We checked that finite temperature has
only a small quantitative effect on our reported fidelities. 
In order to keep our numerically exact simulations feasible we
had to restrict the total number of qubits to ten, and hence the size
of the bath to only two physical qubits when considering two encoded
4-qubit DFS qubits. For consistency we also used two bath qubits in our single
encoded qubit simulations. However, we adjusted the strength of the
coupling to the bath to account for this, using a scaling relation we
found by testing
bath sizes in the range $2\leq N_n \leq 5$.
While we are fully aware of the potential problems associated with using a small
bath at long time scales, we expect our small bath simulations to be reliable, as we are concerned only with short timescales,
relative to the magnitudes of $J$ and $\beta$, so that possible
coherence recurrences due to non-Markovian effects are well out of reach.

\section{Complete simulation results}
\label{app:complete}

Figures~\ref{fullmemory}-\ref{cphase} present our complete gate
fidelity results for the 4-qubit DFS code subject to free evolution 
or CDD, with the $\beta$ and $J$ parameters each varying over the range from 1Hz to 1MHz. The pulse
interval $\tau_0$ is fixed at 1ns, and the pulse width assumes the
values 0, 1ps, and 1ns. In all cases CDD 
leads to a fidelity improvement of many orders of magnitude relative to free evolution. The plots
surrounded by dashed line boxes are the ones shown in
Fig.~1 in the main text.

\section{G\lowercase{a}A\lowercase{s} and S\lowercase{i} parameters}
\label{app:table}

We picked the ranges of our $\beta$ and $J$ parameters shown in
Figs.~\ref{fullmemory}-\ref{cphase} to represent
realistic quantum dot systems -- see Table~1. The $J$ parameters in
this table were estimated directly from the contact
hyperfine interaction between a quantum dot electron and nuclear
spins.  For GaAs the hyperfine constant is $A=90\mu$eV \cite{Coish:06}.
For polarized nuclei, the interaction strength is reduced by the
number of nuclei $N_n$, typically in the range $10^5$ to $10^6$ per
quantum dot, giving $J=A/N_n$. For unpolarized nuclei, the interaction strength is
reduced by $\sqrt{N_n}$ \cite{Coish:06}.  For Si (silicon), 
the contact hyperfine strength $A$ is computed from $A
= \frac{8\pi}{3} \mu_{^{29}Si} \mu_B r \nu \eta$, where $\mu_j$ is the
nuclear or electron (Bohr) magnetic moment, $r$ is the
fraction of non-zero spin $^{29}$Si nuclei, 
$\nu$ is the Si nuclear number density, and $\eta$ is the
concentration of the electron Bloch wavefunction near the Si
nuclei \cite{Hale:59}. For natural abundance Si we obtained $A=60$neV.
For polarized nuclei, the interaction strength is reduced by the
number $N_n\sim 6\times 10^4$ of $^{29}$Si in a quantum dot.
For unpolarized nuclei the interaction strength is again reduced by
$\sqrt{N_n}$.

For both GaAs and Si the energy scale $\beta$ of the bath is taken to
be $1/T_{2n}$, the inverse of the nuclear dephasing time.  For GaAs,
the nuclear dephasing time has been estimated at $T_{2n}=100\mu$s
\cite{khaetskii:186802,merkulov:205309}, which is the precession time of a nucleus in the
dipole magnetic field generated by neighboring nuclei.  For Si, the
nuclear dephasing time has been measured to be $T_{2n}=5.6m$s \cite{Dementyev:03}, again corresponding
to the evolution time due to the dipole-dipole interaction between a
$^{29}$Si nucleus and neighboring $^{29}$Si nuclei.\\

\begin{tabular}{lll}
\hline
Table 1 &  &  \\ \hline\hline
system & \multicolumn{1}{|l}{$J$} & $\beta $ \\ \hline
unpolarized GaAs & \multicolumn{1}{|l}{$100$MHz} & $10$kHz \\ 
polarized GaAs & \multicolumn{1}{|l}{$1$MHz} & $10$kHz \\ 
unpolarized Si & \multicolumn{1}{|l}{$400$kHz} & $180$Hz \\ 
polarized Si & \multicolumn{1}{|l}{$1.5$kHz} & $180$Hz\\
\label{tab1}
\end{tabular}

\begin{figure*}[tp]
\includegraphics[width=6.5in]{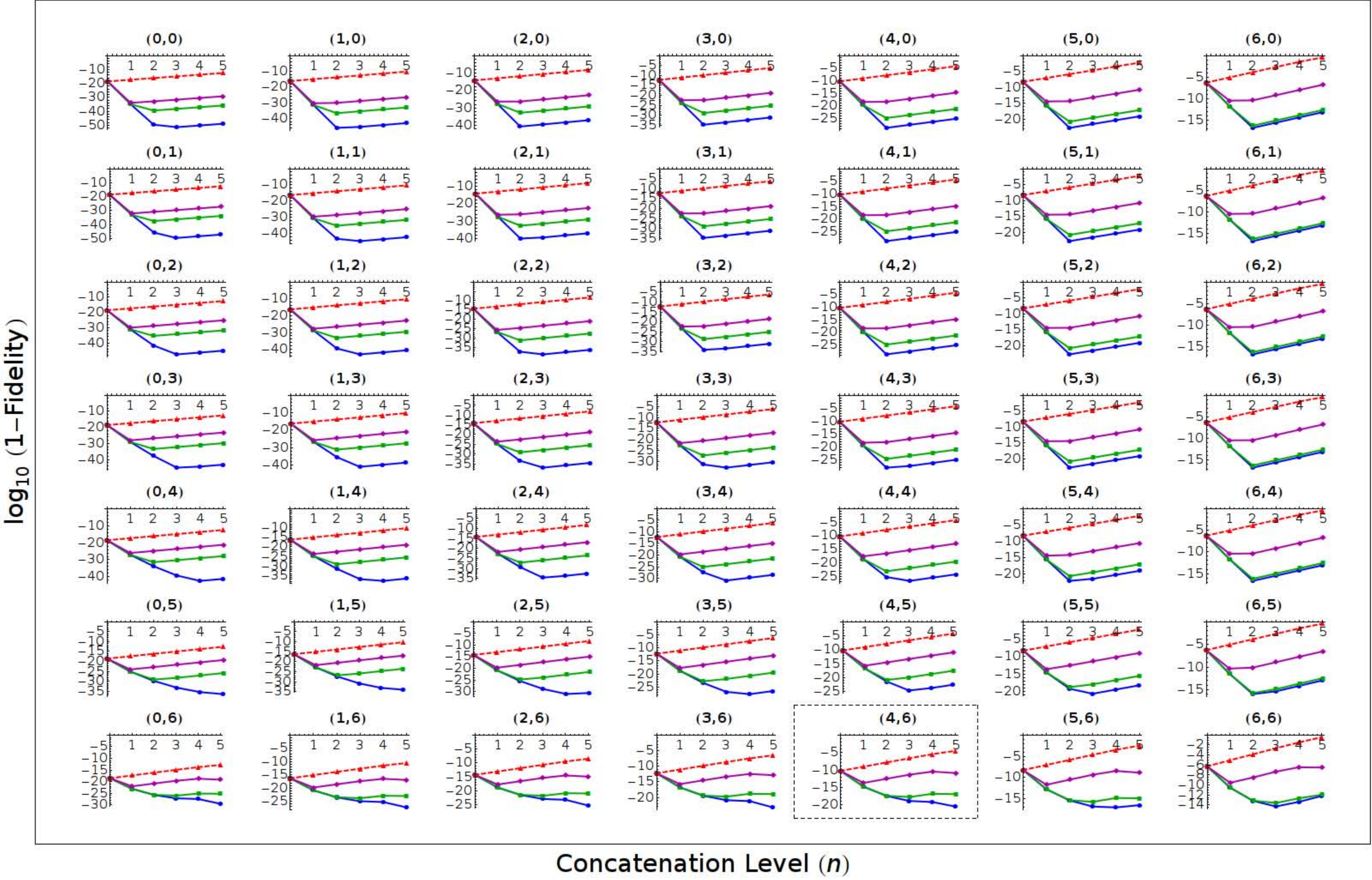}
\caption{CDD quantum memory in the 4-qubit DFS encoding with pulse interval $%
\protect\tau_0 =10^{-9}s$, and pulse widths $\protect\delta =0$s, $\protect%
\delta =10^{-12}$s, and $\protect\delta =10^{-9}$s corresponding to the
blue, green, and magenta lines, respectively. The dashed-red represents free
evolution without DD for increasing total times
$T=4^{n}\protect\tau_0$.
The plots are labeled by the strength of the coupling parameters $(\log_{10}J,\protect%
\log_{10}\beta )$, in Hz. Thus $J$ increases in strength in multiples
of $10$ from left to right, and $\beta$ similarly increases from top
to bottom.
Notice that the $\log _{10}(1-\text{Fidelity}%
) $ ranges change between plots.}
\label{fullmemory}
\end{figure*}

\begin{figure*}[tp]
\includegraphics[width=6.5in]{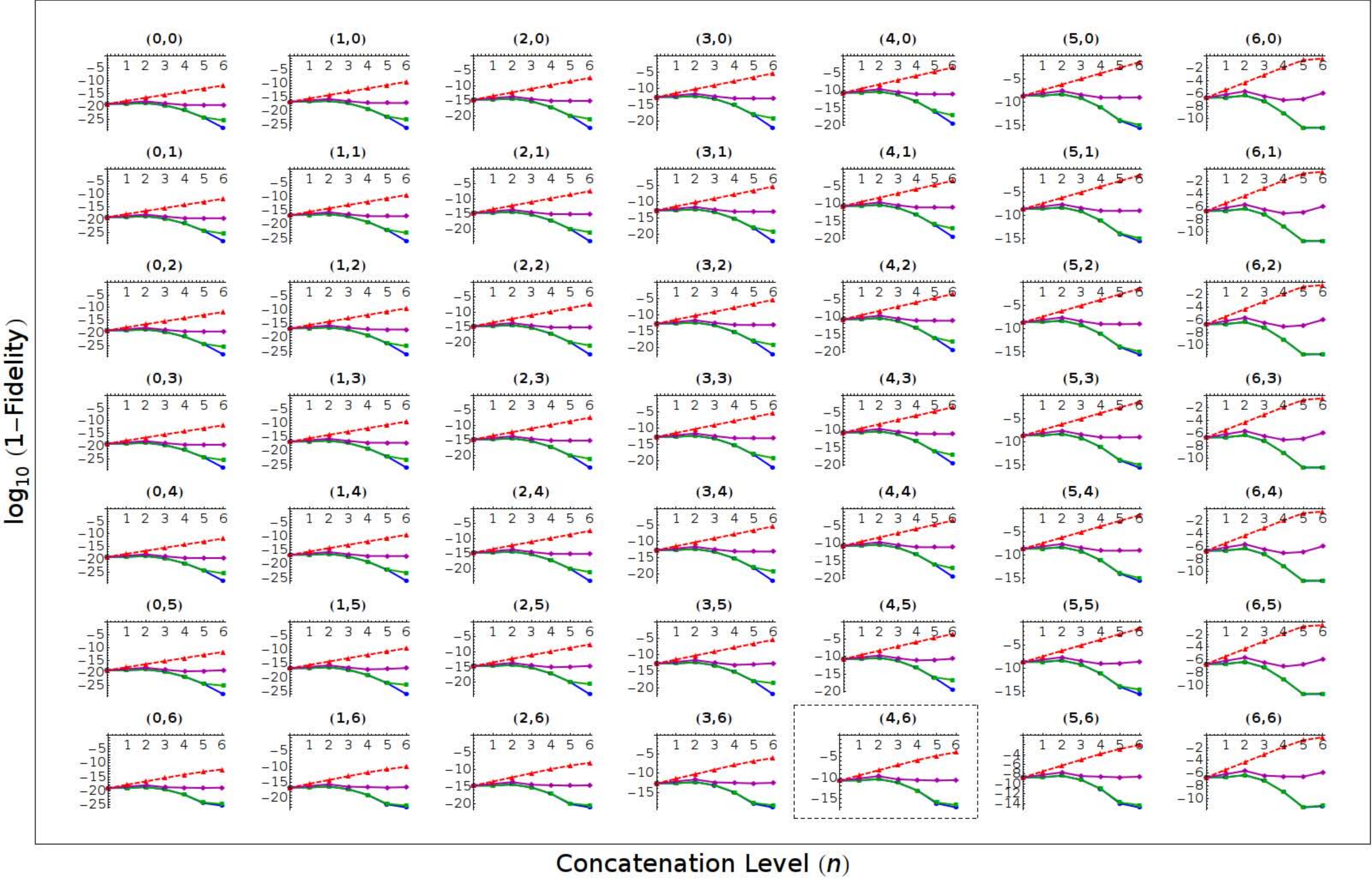}
\caption{Same as Figure~\ref{fullmemory}, for the $\protect\pi /8$ gate.}
\label{fullpi8}
\end{figure*}

\begin{figure*}[tp]
\includegraphics[width=3.2in]{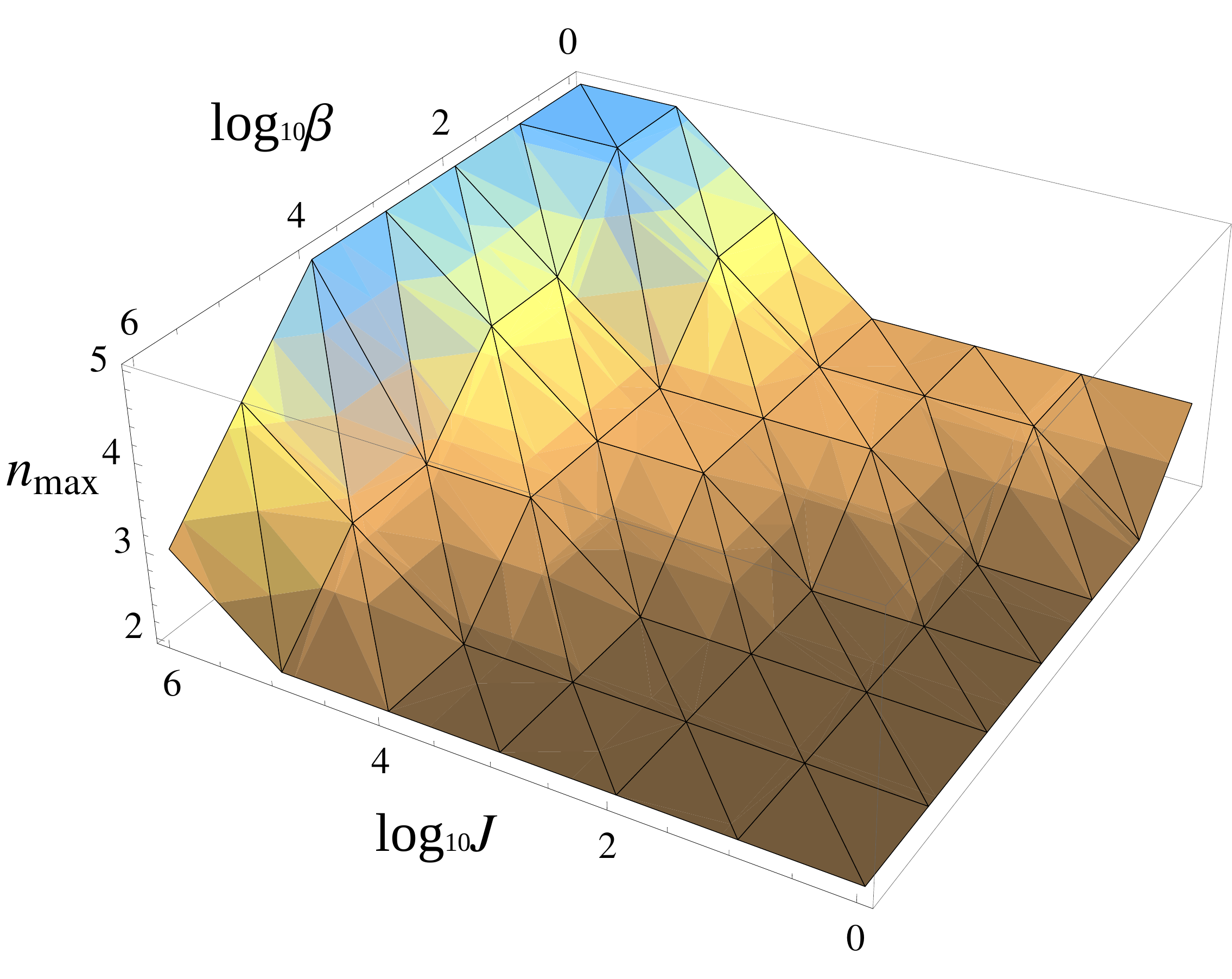}
  \hskip 3mm
\includegraphics[width=3.2in]{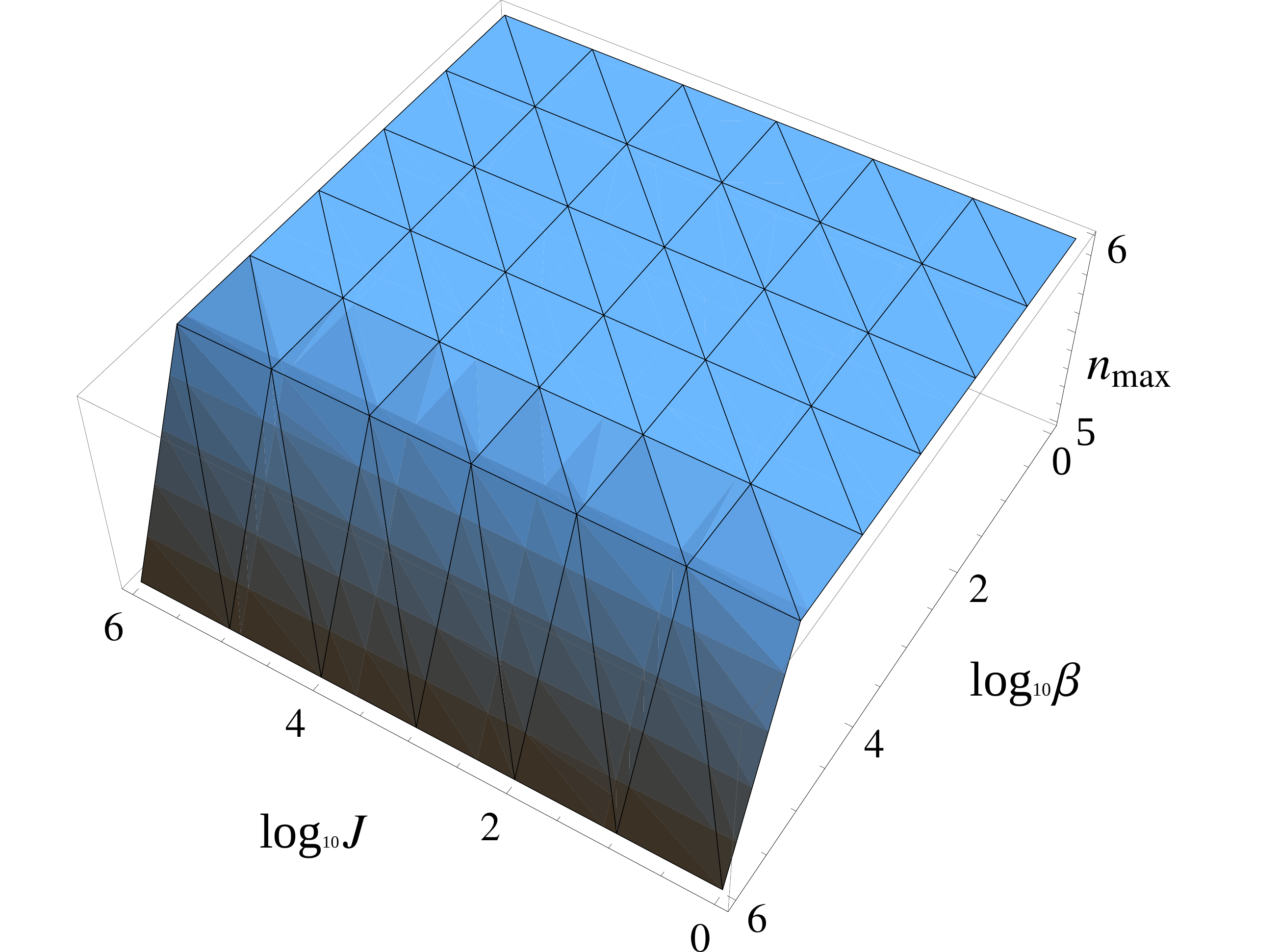}
\caption{Top: Turning point of the memory fidelity shown in Figure~\ref{fullmemory}, i.e., the maximum
  concatenation level before fidelity decreases, as a function of $\log_{10}J$
  and $\log_{10}\beta$, in Hz. Bottom: Same for the $\pi/8$ gate fidelity shown in Figure~\ref{fullpi8}.}
\label{memory-nmax}
\end{figure*}

\begin{figure*}[tp]
\includegraphics[width=6.5in]{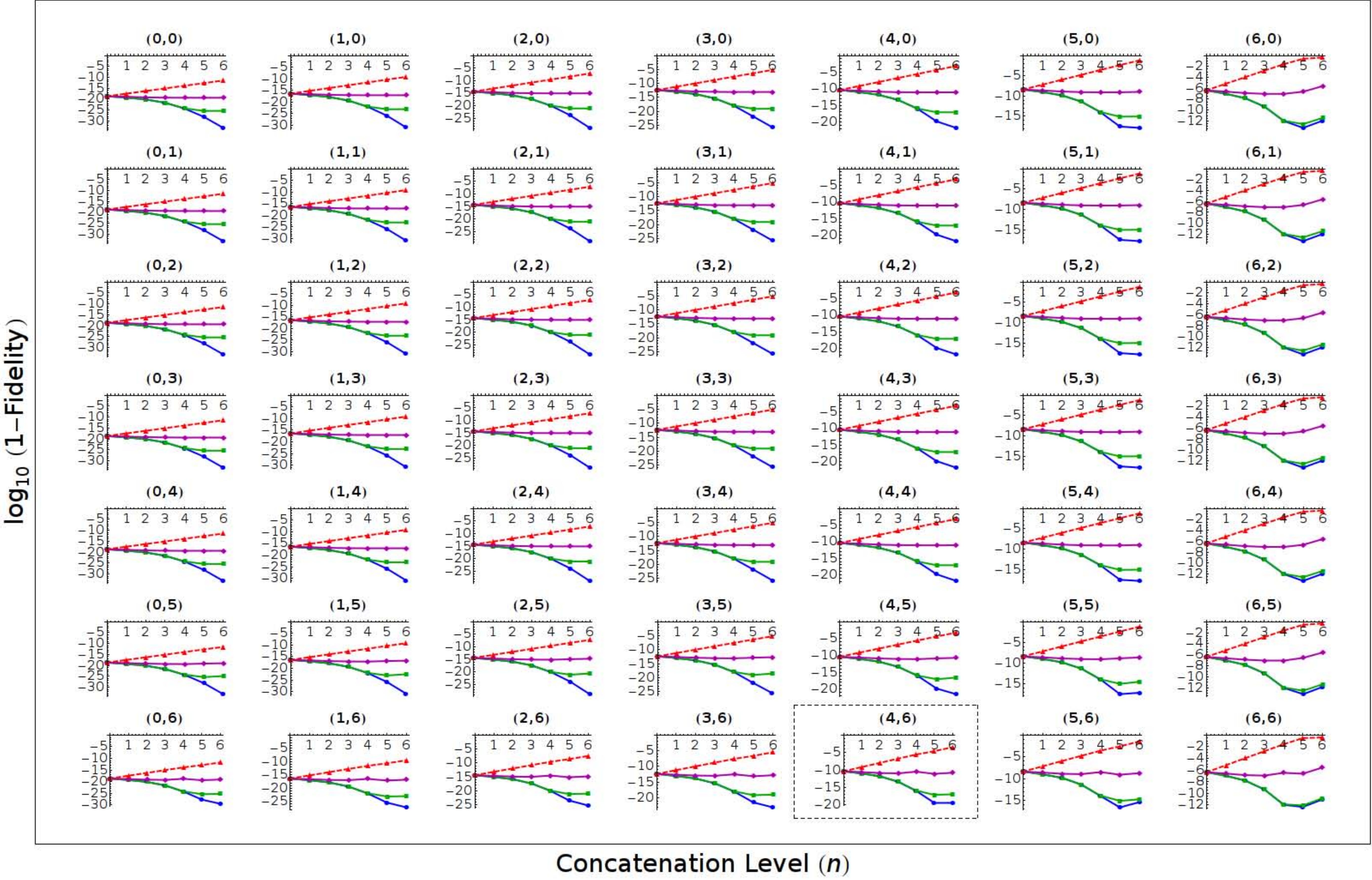}
\caption{Same as Figure~\ref{fullmemory}, for the Hadamard gate.}
\label{fullhadamard}
\end{figure*}

\begin{figure}[tph]
\includegraphics[width=3.2in]{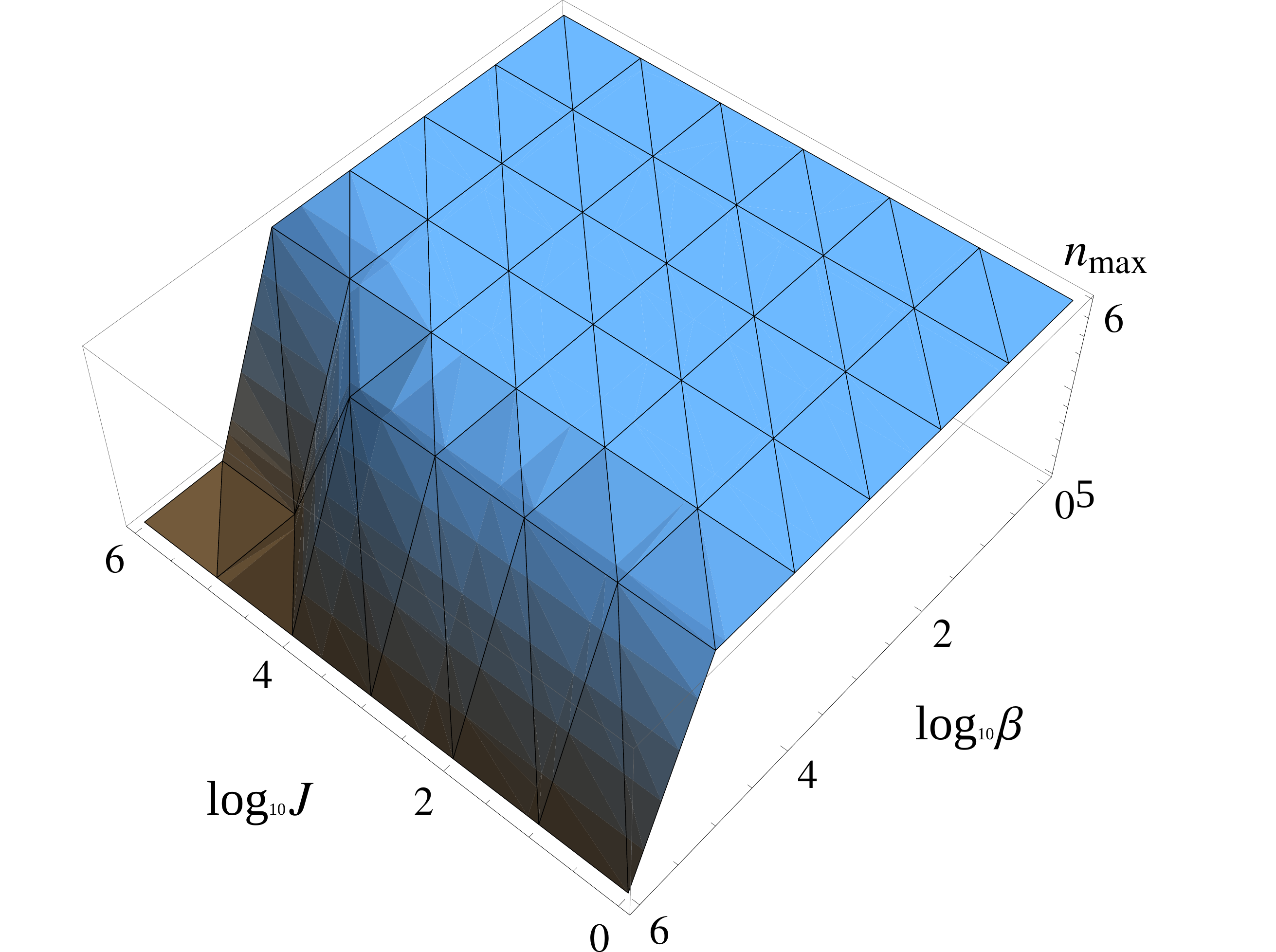}
\caption{Turning point of the Hadamard gate fidelity shown in Figure~\ref{fullhadamard}, i.e., the maximum
  concatenation level before fidelity decreases, as a function of $\log_{10}J$
  and $\log_{10}\beta$, in Hz. Note that slightly better performance is obtained for
  the $\pi/8$ gate. This is due to the fact that it takes only a
  single elementary Heisenberg exchange operation to implement, while
  the Hadamard gate takes two such operations, i.e., double the time.}
\label{Had-nmax}
\end{figure}

\begin{figure*}[tp]
\includegraphics[width=6.5in]{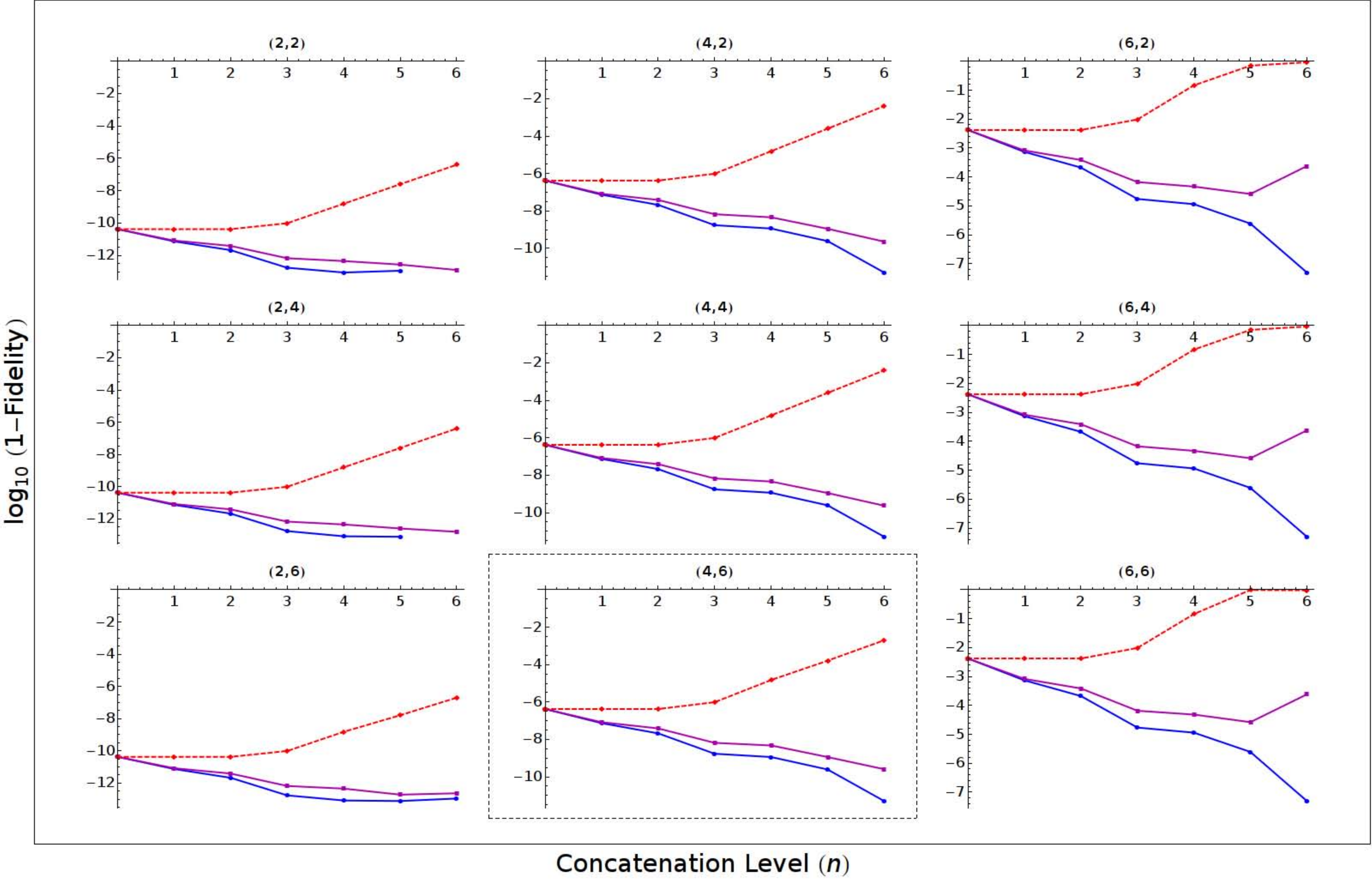}
\caption{Same as Figure~\ref{fullmemory}, for the controlled-phase
  gate.}
\label{cphase}
\end{figure*}

\end{widetext}

\end{document}